\documentclass[12pt]{article}%
\usepackage[normalem]{ulem}
\usepackage{amsfonts}
\usepackage{bm}
\usepackage{cite}
\usepackage{mathrsfs}
\usepackage{amsmath}
\usepackage{slashed}
\usepackage{graphicx} %
\usepackage{hyperref}
\usepackage{verbatim}
\usepackage{cancel}
\usepackage{xcolor}
\usepackage{enumerate}
\usepackage{amsfonts}
\usepackage{amssymb}
\usepackage{authblk}
\usepackage[hang,flushmargin]{footmisc}
\usepackage{lipsum}
\usepackage[font=footnotesize]{caption}
\usepackage{soul}
\usepackage[utf8]{} 
\usepackage{empheq}

\csname @addtoreset\endcsname{equation}{section}
 \newcommand{\badat}{\begin{alignedat}}
 \newcommand{\eadat}{\end{alignedat}}

\newcommand{\eq}{\begin{equation}}
\newcommand{\feq}{\end{equation}}
\newcommand{\eqn}{\begin{eqnarray}}
\newcommand{\feqn}{\end{eqnarray}}

\def\be{\begin{eqnarray}}
\def\ee{\end{eqnarray}}
\def\beann{\begin{eqnarray*}}
\def\eeann{\end{eqnarray*}}
\def\beq{\begin{equation}}
\def\eeq{\end{equation}}
\def\ba{\begin{array}}
\def\ea{\end{array}}
\def\ben{\begin{enumerate}}
\def\een{\end{enumerate}}
\def\bea{\begin{eqnarray}}
\def\eea{\end{eqnarray}}

\def\5{\bar }
\def\6{\partial }
\def\7{\hat }
\def\4{\tilde }
\def\vp{\varphi}

\def\cA{\mathcal{A}}
\def\cB{\mathcal{B}}

\def\cE{\mathcal{E}}
\def\cF{\mathcal{F}}

\def\cK{\mathcal{K}}
\def\cL{\mathcal{L}}

\def\cQ{\mathcal{Q}}

\def\cU{\mathcal{U}}
\def\cV{\mathcal{V}}

\def\cX{\mathcal{X}}

\renewcommand{\d}{\partial}

\renewcommand{\tilde}{\widetilde}
\renewcommand{\hat}{\widehat}

\begin{document}

\title{\vspace{-70pt} \Large{\sc New boundary conditions in \\Einstein-scalar gravity in three dimensions }\vspace{10pt}}
\author[a]{\normalsize{Andr\'es Anabal\'on}\footnote{\href{mailto:
anabalo@gmail.com}{
anabalo@gmail.com}}}
\author[b]{\normalsize{Hern\'an A. Gonz\'alez}\footnote{\href{mailto:hernan.gonzalez@uai.cl}{hernan.gonzalez@uss.cl}}}
\author[c]{\normalsize{An\'ibal Neira-Gallegos}\footnote{\href{mailto:aneira2017@udec.cl}{aneira2017@udec.cl}}}
\author[c]{\normalsize{Julio Oliva}\footnote{\href{mailto:juoliva@udec.cl}{juoliva@udec.cl}}}

\affil[a]{\footnotesize\textit{Instituto de F\'isica Te\'orica, UNESP-Universidade Estadual Paulista, R. Dr. Bento T. Ferraz 271, Bl. II, Sao Paulo 01140-070, SP, Brazil.}}
\affil[b]{\footnotesize\textit{Universidad San Sebasti\'an, Avenida del C\'ondor 720, Santiago, Chile.}}
\affil[c]{\footnotesize\textit{Departamento de F\'isica, Universidad de Concepci\'on, Casilla, 160-C, Concepci\'on, Chile.}}

\date{}

\maketitle
\thispagestyle{empty}
\begin{abstract}
We analyze the backreaction of a class of scalar field self-interactions with the possibility of evolving from an AdS vacuum to a fixed point where the scalar field potential vanishes. Exact solutions which interpolate between these regions, ranging from  stationary black hole to dynamical spacetimes are constructed. Their surface charges are finite but non-integrable. We study the properties of these charges on the solutions. In particular, we show that the integrable part of the charges provides a realization of the conformal algebra by means of a modification of the Dirac bracket proposed by Barnich and Troessaert. The latter construction allows for a field dependent central extension, whose value tends to the Brown-Henneaux central charge at late times.  

\end{abstract}

\newpage

\begin{small}
{\addtolength{\parskip}{-2pt}
 \tableofcontents}
\end{small}
\thispagestyle{empty}
\newpage

\section{Introduction}

The idea of holography has its more concrete realization in AdS/CFT, which allows to describe a strongly coupled conformal field theory in terms of the physics in an asymptotically AdS background. Indeed, it is believed that this idea can be generalized to the Gauge/Gravity duality, where any observable of a gravitational theory can be described in terms of a gauge theory and vice versa. A step forward in this direction would be to study asymptotically flat holography by means of its relation with  an asymptotically AdS spacetime. 

From another point of view, the identification of infinite-dimensional symmetries at the boundary of the spacetime have served as an important tool to unfold the holographic properties of quantum gravity. The prime example is the analysis made by Brown and Henneaux, where the 2D conformal symmetry emerging on asymptotically AdS$_3$ spaces \cite{Brown:1986nw} has shed light on the asymptotic growth of the black hole microstates \cite{Strominger:1997eq}. More recently, the prominent role played by the infinite-dimensional BMS group arising in asymptotically flat spaces has provided new connections with infrared physics of gauge theories (see \cite{Strominger:2017zoo} for a review on the subject).  

Here, we would like deepen our understanding of the holographic properties of gravity by identifying new realizations of the asymptotic symmetry group (ASG) in three-dimensional spacetimes with non-standard fall-off conditions. This is achieved by focusing on gravity minimally coupled to a scalar field with a family of interacting potentials  given by
\be
V(\Phi)= c_1 e^{-\Phi}+ c_2 e^{-\tfrac{3}{2}\Phi}+ \cdots\,,
\label{family}
\ee
with $c_i$ fixed constants and the ellipsis representing faster decaying terms. The above type of potentials are of physical interest as they appear in string theory compactifications \cite{Deger:2014ofa, Deger:2019jtl} and also they can be inspired by dimensional reductions where the scalar field represents the size of the extra dimensions \cite{Carroll:2001ih}. In the present context, they 
allow us to study the asymptotic behavior of geometries whose Riemann tensor approaches to zero in the region far away from localized sources
\be
\label{ALF}
R^{\alpha \beta}_{\,\,\,\,\,\,\gamma \delta} \to 0\,.
\ee
Note that this condition does not necessarily imply that the metric resembles Minkowski space in the asymptotic region (see \cite{Oliva:2009ip} for an example where the leading components of the metric diverge at the boundary). We will refer to configurations satisfying \eqref{ALF} as \emph{asymptotically locally flat spaces}  \footnote{ In order to be consistent, the notion of asymptotic flatness considered here has to be referred to a specific frame and in a neighbourhood of the conformal infinity. In this work,  we mainly focus on the \emph{Einstein frame}. Nevertheless, the reader may refer to the discussion in subsection \ref{cjf} for a brief analysis of the asymptotic structure in the Jordan frame.}. Black hole solutions exhibiting this kind of asymptotics have been found in Einstein-scalar theories with an exponential potential that is unbounded from below in \cite{Chan:1994qa,Chan:1995wj,Charmousis:2001nq,Charmousis:2009xr}.

In this article we report the existence of an infinite-dimensional asymptotic symmetry\footnote{Asymptotic symmetry groups relevant to asymptotically flat spaces in three-dimensional gravity coupled to matter fields, have been studied in \cite{Spindel:2018cgm,Detournay:2018cbf}.} isomorphic to the Virasoro group. The surface charges associated to these symmetries are however non-integrable in the space of solutions. This is commonly expected in spacetimes displaying radiative degrees of freedom at the boundary, like in four-dimensional asymptotically flat Einstein gravity \cite{Barnich:2011mi,Barnich:2013axa} or in AdS gravity with enhanced boundary conditions \cite{Compere:2019bua,Compere:2020lrt}. 
Due to the lack of local graviton excitations in three dimensions, one can relate the existence of the non-integrable charges to the presence of a boundary scalar field that propagates the only degree of freedom present in the model. 

In order to define brackets for generators realizing canonically this infinite-dimensional symmetry, we employ the definition proposed by Barnich and Troessaert for charges exhibiting non-integrable contributions \cite{Barnich:2011mi}. Interestingly enough, the Barnich-Troessaert bracket leads us to an algebra where a new generator representing the scalar radiation in the system can be defined. Furthermore, this generator becomes a center of the algebra at very late times, therefore, it is recognized as the central charge of the Virasoro algebra.  

In three dimensions, a method to determine all axially symmetric and stationary solutions to the Einstein-scalar system has been developed in \cite{Aparicio:2012yq}. The results presented in this article partially extend the previous algorithm by constructing novel non-circularly symmetric solutions with enhanced asymptotics. 

The organization of the paper is as follows. In the next section \ref{s2}, we describe the specific model that allows to construct solutions and provide a characterization of asymptotically locally flat solutions endowed with scalar fields exhibiting logarithmic profile at infinity.  

In section \ref{s3}, we find an infinite-dimensional enhancement of the asymptotic symmetries defined in terms of a Weyl transformed Bondi gauge. We then show, in section \ref{s4}, that the phase space of solutions spanned by the boundary conditions are consistent with the equation of motions near the boundary for the family of self-interacting potentials \eqref{family}. Moreover, the corresponding functional variations of the charges turn out to be finite  even though they  receive divergent contributions  from gravity and the scalar matter sector. 

The non-integrability of the charges is explicitly presented in \ref{exact}, by describing two types of exact solutions. In \ref{exact:1} a stationary black hole is constructed. Although its mass is non-integrable, it is shown to be consistent with the first law of thermodynamics. The self-interacting potential of the theory has a negative minimum that corresponds to a locally AdS$_3$ solution. We prove that this type of space-time emerges as the near horizon geometry in the region of the parameters where the solution becomes extremal.  The second family of solutions are displayed in section \ref{exact:2}, where we construct dynamical spacetimes containing an infinite set of boundary degrees of freedom encoded in a leading component of the scalar field. The time dependence of the geometry exhibits an evolution such that at very late times it decays into a locally AdS$_{3}$ space. This is possible due to a minimum in the self-interacting potential. In section \ref{alg}, we show the evolution of this geometry in terms of non-integrable charges and its corresponding algebra. We finalize this article in section \ref{s7} describing the main lessons learned and sketching some that will be addressed elsewhere.

\section{The model}
\label{s2}
We are interested in the study of the asymptotic structure of asymptotically locally flat spaces in the minimally coupled Einstein-scalar theory. The action principle is given by
\begin{equation}
\label{eq:action}
    S_0[g_{\mu \nu}, \Phi ]=\frac{1}{16\pi G}\int d^3x \sqrt{-g} \left[R -\frac{1}{2} \nabla_\mu \Phi \nabla^\mu \Phi -V(\Phi) \right]\,
\end{equation}
where the self-interacting potential $V(\Phi)$ is unspecified at this point. The field equations that follow from the extremization of \eqref{eq:action} produce 
\begin{equation}
\begin{split}
\label{EOMphi}
E^{\mu}_\nu&:= G^{\mu}_\nu - \frac{1}{2}\left[\nabla^\mu \Phi \nabla_\nu \Phi - \delta^{\mu}_{ \nu} \left(\frac{1}{2} \nabla_\alpha \Phi \nabla^\alpha \Phi +V(\Phi)\right) \right]=0\,,\\
E_{\Phi}&:= \Box \Phi -V'(\Phi)=0\,.\\
\end{split}
\end{equation}
We would like to characterize asymptotically locally flat solutions for the action \eqref{eq:action}, for specific potentials $V(\Phi)$.As an example of the main features described by these spaces, we will concentrate in this section on the family of solutions obtained by Mann and Chan \cite{ Chan:1994qa,Chan:1995wj}\footnote{ The emergence of these solutions in the context of low-energy string compactifications and its relation to holography has been studied in \cite{Kanitscheider:2008kd}.}. This configuration is the extremum of  \eqref{eq:action} with an exponential potential\footnote{$D$-dimensional solutions to the Einstein-scalar theory with this type of potential have been found in  \cite{Charmousis:2001nq}.}
\begin{equation}
\label{CM-V}
V(\Phi)=- \frac{(2-\alpha)}{g^2}\, e^{-\sqrt{2\alpha} \Phi}\,,
\end{equation}
with $\alpha>0$ and $g$ a fixed constant with dimensions of length. The circularly symmetric solution reads
\begin{equation}
\label{CMsol}
    ds^2= -\left( \frac{r^2}{g^2}- M r^\alpha \right)dt^2 + \left(\frac{r}{L} \right)^{2\alpha} \frac{dr^2 }{\frac{r^2}{g^2}-M r^\alpha} + r^2 d\varphi^2 
\, , \quad
    \Phi(r)=\sqrt{2\alpha}\log\left(\frac{r}{L}\right)\,.
\end{equation}
The above configuration possesses two integration constants, $M$ and $L$. When $M>0$ and $0<\alpha<2$ the metric represents a black hole  whose curvature singularity is being hidden by the event horizon located at $r_0=(g^2 M)^{\tfrac{1}{2-\alpha}}$.

This geometry is an example of an asymptotically locally flat spacetime as the Ricci tensor\footnote{In three dimensions, the Riemann tensor can be completely expressed in terms of the Ricci tensor.} behaves like $R^{\mu}_{\,\,\nu}=O(r^{-2\alpha})$
i.e., vanishes at $r\to \infty$. Despite this feature, its conformal boundary shares similarities with AdS$_3$ spaces. In order to make this statement more precise, let us perform the following change of coordinates
\begin{equation}
    r=\frac{\rho^{q+1}}{L^q}\,,\quad t=\frac{\tau}{q+1}\,,  \quad  \alpha=\frac{q}{q+1}\,.
\end{equation}
Under this transformation,  solution \eqref{CMsol} becomes
\begin{equation}
\label{confCM}
    ds^2=\left(\frac{\rho}{L} \right)^{2q} \left[ -\left( \frac{\rho^2}{l^2} -\frac{m}{\rho^{q}} \right) d\tau^2 + \frac{d\rho^2}{\frac{\rho^2}{l^2}  - \frac{m}{\rho^{q}} }+ \rho^2 d\varphi^2  \right]\,,\quad
    \Phi(\rho)=\sqrt{2q(q+1)} \log\left(\frac{\rho}{L} \right)\,.
\end{equation}
where we have redefined $l$ and $m$ by $l=(q+1) g$ and $m=\tfrac{ L^{\alpha +q}}{(q+1)^2}M $. It is then interesting to note that for $q \geq -1$ the metric becomes conformal factor times an asymptotically AdS$_{3}$ spacetime in the sense of Brown-Henneaux \cite{Brown:1986nw}.
In fact, we can define a regular metric at infinity by taking $\rho=\frac{1}{\Omega}$. Then at $\Omega=0$, the boundary metric is defined by
\begin{equation}
  \lim_{\Omega\to 0}L^{2q}\,\Omega^{2(q+1)} ds^2=-\frac{d{\tau}^2}{l^2} + d\vp^2\,,
\end{equation}
which means that the full two-dimensional conformal group generates transformations at the boundary.

It is important to highlight that the process of bringing the boundary of the spacetime to a finite region is controlled by the real number $q$, that has been initially defined as one of the potential's coupling constant. In this sense, this is similar to the notion of conformal infinity constructed in \cite{Bonga:2020fhx} for cosmological spacetimes, where $q$ represents a deceleration parameter of the fluid's equation of state. 

In order to explore the properties of the conformal boundary associated to spaces \eqref{confCM}, we shall consider, for the rest of this article, geometries equipped with $q=1$ asymptotics. In the appendix \ref{Exactq}, we comment on a exact solution with arbitrary $q$.

\section{Asymptotic symmetries}
\label{s3}
We study the asymptotic group of a broader set of  solutions displaying fall-off conditions incorporating  \eqref{confCM} with $q=1$. In order to analyze radiative boundary conditions and also compare with similar findings of  standard asymptotically flat spacetimes at null infinity \cite{Barnich:2010eb,Barnich:2012aw}, we make use of a Weyl transformed version of the so-called Bondi-Metzner-Sachs (BMS) gauge \cite{Bondi:1962px,Sachs:1962zza}.  

Before delving into the computations, it is important to mention that one can expect that the presence of a non-zero scalar field in the bulk to break the isometries of a given background. In this sense, the behavior of these symmetries has been explored in a holographic setting by considering the so-called ``generalized conformal structure'' \cite{Kanitscheider:2008kd}. Here, we are going to provide a complementary view of this structure from a purely gravitational perspective by analyzing the asymptotic symmetries preserved at the boundary $\rho=\infty$. As we will see below, studying the symmetries that only preserve asymptotic configurations reveals that they are, in fact, infinitely enhanced.

We focus on a family of spacetimes characterized by  
\begin{equation}
\label{CBondi}
    ds^2=e^{\Phi} \left[ \mathcal{V}(u,\rho,\varphi) du^2 - 2 du d\rho + 2\,\mathcal{U}(u,\rho,\varphi) du d\varphi + \rho^2 d\varphi^2 \right]\,, \quad \Phi=\Phi(u,\rho,\varphi)\,.
\end{equation}
In this gauge choice $u$ is a light-like coordinate, $\rho$ is a radial direction and $\varphi$ is an angle orthogonal to the null rays $u={\rm const}$.  An  important feature of this gauge is that we have introduced a conformal factor that explicitly depend on the scalar field $\Phi$. The metric \eqref{CBondi} leads to the black hole spacetime \eqref{confCM} for $q=1$ when $\mathcal{V}=-\frac{\rho^2}{l^2}+\frac{m}{\rho}$ and $\cU=0$. The symmetries preserving these gauge conditions must fulfill
\begin{equation}
\label{gfc}
\mathcal{L}_\xi g_{\rho\rho}=\mathcal{L}_\xi g_{\rho\varphi}=0\,, \quad \mathcal{L}_\xi g_{\varphi\varphi}= \mathcal{L}_\xi \Phi\, g_{\varphi\varphi}\,
\end{equation} 
which lead to the definition of the components of the gauge-preserving vector field
\begin{equation}
\label{xiasym}
    \xi^u=f\,, \quad 
    \xi^{\varphi}=Y-\frac{1}{\rho}\partial_\varphi f\,, \quad
    \xi^\rho=-\rho \partial_\varphi Y + \partial_{\varphi}^2 f -\frac{1}{\rho} \partial_{\varphi} f\, \mathcal{U}\,,
\end{equation} 
where $\partial_\rho f=\partial_\rho Y=0$. Notice that the vector field $\xi$ includes an explicit dependence on the metric function $\cU$. 

 To fully characterize the fields at infinity, it is necessary to provide a suitable radial decay of the fields $\cV$, $\cU$ and $\Phi$ consistent with known solutions like \eqref{confCM}. Here we will be interested in the preservation of the following fall-off conditions 
\begin{equation}
\label{asym}
\mathcal{V}=-\frac{\rho^2}{l^2}+\cV_0+O(\rho^{-1})\,, \quad \mathcal{U}=\cU_0+O(\rho^{-1})\,, \quad 
\Phi=2\log\left(\frac{\rho}{l}\right) -2\theta+O(\rho^{-1})\,,
\end{equation} 
where $\cV_0$, $\cU_0$ and $\theta$ are arbitrary functions of $u$ and $\varphi$, while $l$ stands for a fixed coupling constant with dimensions of length. It is interesting to notice that $l$ corresponds to the AdS radius of asymptotic region of the conformally related spacetime, whose line element is $e^{-\Phi} ds^2$. 

By acting with  \eqref{xiasym} on the asymptotic expansions, we find that the vector field is further restricted by conditions \eqref{asym}. This implies that the components $f$ and $Y$ are further restricted by the equations
\begin{equation}
\label{eq:fY}
\dot{Y} =\frac{1}{l^2}f'\,, \quad \dot{f} =Y'\,,
\end{equation} 
where $\d_\vp$ is represented by primes $'$ and `dots' are time derivative, $\d_u$. 

For $l \neq 0$, these functions comprise a representation of two copies of the Witt algebra. This can be proved by using a modified version of the Lie bracket \cite{Barnich:2010eb}
\begin{equation}
[\xi(s_1,g),\xi(s_2,g)]=\mathcal{L}_{\xi(s_1,g)}\xi(s_2,g) -\delta_{\xi(s_1,g)} \xi(s_2,g)+ \delta_{\xi(s_2,g)} \xi(s_1,g)\,.
\end{equation}
where $s_i$ stands for the pair $(f_i,Y_i)$, and the last two terms have been added due to the explicit dependence of the Killing vectors \eqref{xiasym} on the metric $g_{\mu \nu}$, as mentioned above. These terms correspond to an infinitesimal change of $\xi(s_2,g)$ due to a diffeomorphism $\xi_1$. Doing so one gets, 
\begin{equation}
[\xi(s_1,g),\xi(s_2,g)]=\xi([s_1,s_2],g)\,.
\end{equation}
where the bracket $[s_1,s_2]=(f_{[1,2]},Y_{[1,2]})$ is a two-dimensional representation of the conformal algebra with  
\begin{equation}
\label{Lie12}
f_{[1,2]}=f_1 Y_2'-f_1' Y_2 +Y_1 f_2'-Y_1'f_2\,, \quad
Y_{[1,2]}=Y_1 Y_2'-Y_1' Y_2 +\frac{1}{l^2}(f_1 f_2'-f_1'f_2).
\end{equation}
It is worth mentioning that the coordinate system chosen in this section admits an smooth $l\to \infty$ limit of the Killing vectors and asymptotic conditions. By dropping $l^{-2}$ terms in \eqref{eq:fY} and \eqref{Lie12}, the resulting algebra generates the BMS$_3$ group \cite{Ashtekar:1996cd,Barnich:2006av}. 

\section{Phase space and charges}
\label{s4}

\subsection{Asymptotic solutions} \label{ASYMP}

We now analyze the space of solutions  of the Einstein-scalar system \eqref{EOMphi} in coordinates \eqref{CBondi} subjected to boundary conditions \eqref{asym}. One of the important advantages of the conformal BMS gauge discussed before, is that one can integrate the radial dependence of the system without making use of the specific functional form of the self-interacting potential  $V(\Phi)$. This property will help us to find the asymptotic radial expansion associated to the fields $\Phi$, $\mathcal{U}$ and $\mathcal{V}$. Once they have been determined we will be able to provide an asymptotic form for $V(\Phi)$ consistent with our boundary conditions.

We start by considering the equation $E^u_\rho=0$, which gives an equation for the scalar field
\begin{equation}
\label{sca}
(\partial_\rho \Phi)^2 +2\partial^2_\rho \Phi=0
\end{equation}
whose exact solution can be written as
\begin{equation}
\Phi(u,\rho,\phi)=2\left( \log\left[\frac{\rho}{l} +\theta_1(u,\varphi) \right]- \theta(u,\varphi)\right)\,.
\end{equation}
Field equations $E^u_u-E^\rho_\rho=E^{u}_{\varphi}=E^{\varphi}_{\rho}=0$ determine the radial dependence on $\mathcal{U}$ from
\begin{equation}
\rho(\rho+l \theta_1)\partial_\rho^2 \mathcal{U}- l \theta_1\partial_\rho \mathcal{U} -2(\mathcal{U}-\mathcal{U}_0)=0\,, \quad \mathcal{U}_0\equiv l( \theta_1 \theta' - \theta_1')\,.
\end{equation}
The latter can be solved in a expansion series consistent with the asymptotic behavior at infinity \eqref{asym},
\begin{equation}
\label{U}
\mathcal{U}=\mathcal{U}_0+\frac{\mathcal{U}_1}{\rho} \sum_{n \geq 0} \frac{3}{n+3} \left(\frac{-l\theta_1}{\rho}\right)^n\,.
\end{equation}
where $\mathcal{U}_{1}=\mathcal{U}_{1}(u,\phi)$ is an arbitrary function of the angles. From $E^\rho_\rho-E^{\varphi}_\varphi=0$ and using the above solution we can find 
\begin{multline}
\rho^2(\rho+l \theta_1)\partial_\rho^2 \cV- l \rho \, \theta_1  \partial_\rho \cV -2 \rho (\cV-\cV_0)=2 ( \cU-\cU_0)\theta' +2(\cU-\cU_0)' \\  + 
(\rho \theta' + \cU_0)\d_\rho \cU + (l\theta_1 + \rho) \d_\rho \cU'\,.
\end{multline}
with $\cV_0=\theta'^2 -\theta''+l(\theta_1 \dot{ \theta} - \dot{\theta}_1)$. The solution to the latter equation can be again expressed in terms of infinite series around $\rho\to \infty$ as
\begin{equation}
\label{V}
\mathcal{\cV}=-\frac{\rho^2}{l^2}+\cV_0+\frac{1}{\rho} \sum_{n \geq 0} \frac{1}{\rho^n} \left(\frac{3 (-l\theta_1)^n \cV_1}{n+3}+ \cF_{n}\right)\,.
\end{equation}
where $\cV_1$ is an arbitrary function of $u$ and $\vp$ while $\cF_n$ are functions completely determined by $\cU_1$, $\theta_1$ and $\theta$. Its particular form is not relevant for the present discussion, except for the fact it vanishes when $\cU_1=\theta_1=0$. 

The only remaining equations are $E^{\rho}_u=0$ and $E^{\rho}_\varphi=0$. They can be solved up to $O(\rho^{-4})$ terms, obtaining an expression for $\theta_1$
\be
\label{th1}
\theta_1=k e^{\theta} + l\dot{\theta}\,.
\ee
where $k$ an integration constant. The sub-leading contributions in the large-$\rho$ expansion imply differential equations involving $\cF_n$, $\cU_1$ and $\theta$. For the purpose of providing the asymptotic solution considered here, it is only necessary to consider the equations arising at $\rho^{-4}$ order. They yield relations for $\cV_1$ and $\cU_1$
\begin{equation}
\begin{split}
&2\dot{\cV}_1-3\cV_1 \dot{\theta} -\frac{3}{l^2}\left( \cU_1' - \cU_1 \theta' \right)=-2(\cV_0 - l\dot{\theta}_1) \Box \theta+ 3 \theta' \Box \theta' + l \theta_1 \Box \dot{\theta}\,,\\ 
&3 \dot{\cU}_1 - 3\cU_1 \dot{\theta} - \cV_1'= l\theta_1\left( 2\theta' \Box \theta - \Box \theta'\right)\,.
\end{split}
\end{equation}
where $\Box \theta= -l^2 \ddot{\theta}+\theta''$ is the two-dimensional d'Alambertian.

Finally, the diagonal components of the field equations $E^u_{u}=E^\rho_{\rho}=E^\varphi_\varphi$ determine the self-interacting potential for large values of the scalar field $\Phi$ to be
\be
\label{familyV}
V(\Phi)=-\frac{6}{l^2} e^{-\Phi}+\frac{4k}{l^2}e^{-\frac{3}{2}\Phi}+ O(e^{-\frac{5}{2}\Phi})\,.
\ee
Notice that the first term corresponds exactly to \eqref{CM-V} for $\alpha=1$, while the following term is a  subleading corrections and thus we must regard $k$ as a fixed coupling constant. In section \eqref{exact}, we will provide exact potentials and their corresponding solutions.

\subsection{Surface charges}
\label{sc}
Surfaces charges associated to \eqref{eq:action} can be computed using the Iyer-Wald method \cite{Iyer:1994ys}. The explicit expressions receive gravitational and scalar contributions and it is given by 
\be
\label{chargesphig0}
\slashed{\delta} Q_\xi = \frac{1}{16 \pi G}\int (k^{\mu \nu}_{g}+k^{\mu \nu}_{\Phi}) \epsilon_{\mu \nu \gamma} dx^{\gamma}= \frac{1}{8 \pi G} \oint \,\left(k^{u\rho}_{g}+k^{u \rho}_{\Phi}\right)\,, 
\ee
where we have defined $\oint\equiv \int^{2\pi}_0 d\varphi $ and $\delta$ stands for an infinitesimal variation in the space of solutions with $h_{\mu \nu}=\delta g_{\mu \nu}$. The explicit expressions for the one-form forms $k_g$ and $k_\Phi$ are
\be
&k^{\mu \nu}_g = 2\sqrt{-g} \left[ \xi^{[\mu} \nabla_\alpha h^{\nu]\alpha} - \xi^{[\mu} \nabla^{\nu]}h - \frac{1}{2}h \nabla^{[\mu}\xi^{\nu]} - \xi_\alpha \nabla^{[\mu} h^{\nu] \alpha} + h^{\alpha [\mu}\nabla_\alpha \xi^{\nu]} \right]\,\label{qq}\\
 &k^{\mu \nu}_\Phi =-2\sqrt{-g}(\xi^{[\mu} \nabla^{\nu]}\Phi \, \delta\Phi)\,.
\ee
Since it is not evident that \eqref{chargesphig0} would be express as an exact $\delta$-variation, we have denoted the  left-hand side of this equation as $\slashed{\delta} Q_\xi$.

The above expressions are evaluated in the asymptotic region $\rho \to \infty$, using the asymptotic Killing vector \eqref{xiasym}.  It is worthwhile mentioning that the Barnich-Brandt formalism \cite{Barnich:2001jy} yields a different expression for the gravitational contribution, however it coincides with the one presented here when using the conformal Bondi gauge \eqref{CBondi} (see \cite{Ruzziconi:2020wrb} for more details on this point).

For the asymptotic conditions previously introduced, when performing the expansion for large $\rho$, both contributions to the charge exhibit linear, quadratic and even cubic divergences. Interestingly enough, these terms cancel out when considering both expression in \eqref{chargesphig0}. 
A similar behavior has been observed in a setup leading to relaxed AdS boundary conditions due to a slowly decaying scalar field at infinity \cite{Henneaux:2002wm,Henneaux:2004zi,Henneaux:2006hk}. These examples, together with the extended asymptotics presented here, highlight the important contribution of the matter fields to render the charges finite in asymptotically locally flat scenarios.

The explicit expression for \eqref{chargesphig0} is generically non-integrable in the space of fields but it can be simplified by using the definition $\cE \equiv e^{-\theta}$. We express the final result as 
\be
\label{chargesphig}
\slashed{\delta} Q_\xi (\cX) =\delta \cQ(s,\cX) +\Theta_s(\cX,\delta \cX)
\ee
with $s=(f,Y)$ and 
\begin{equation}
\label{charge}
\begin{split}
\cQ(s,\cX)&=\frac{1}{16 \pi G} \oint \,\left[(k-l\dot{\cE})\left( f \cV_0 + 2Y\cU_0\right) +\frac{2}{l} f \cE \cV_1+ \frac{3}{l} Y\cE \cU_1\right],\\
\Theta_s(\cX,\delta \cX)&=\frac{1}{16 \pi l G} \oint \, \left(4 f'  \cU_0 \delta \cE + 2 f \cU_0' \delta \cE + l^2 f \cV_0 \delta \dot{\cE}   + f  \cV_1\delta \cE \right)\,.
\end{split}
\end{equation}
where $\cQ(s,\cX)$ is the integrable piece, while $\Theta_s(\cX,\delta \cX)$ is the contribution that cannot be expressed as a local functional of the fields. This definition is of course not unique, as we can always add an arbitrary functional $N_s$ and define $\cQ'=\cQ+N_s$ with $\Theta'=\Theta-\delta N_s$, such that \eqref{chargesphig} is invariant. 

A canonical realization of the asymptotic symmetries needs a definition of an integrable charge $Q_{\xi}$. One way of sorting out this issue  is by restricting our phase space to configurations satisfying $\delta \Theta(\cX,\delta \cX)=0$. An immediate consequence of the latter shows that one cannot provide a functional relation among the fields $\cE$, $\dot{\cE}$ and $\cV_1$
without breaking the enhanced asymptotic symmetry. Hence, in section \ref{alg}, we are going to proceed with a different approach, understanding the canonical properties of the integrable charge $\cQ(s,\cX)$.

\section{Exact solutions}
\label{exact}
 Here we present two classes of exact solutions realizing the asymptotic behavior studied in section \eqref{ASYMP}.

\subsection{Stationary black hole solution}
\label{exact:1}
We present an extension of Mann-Chan solution \eqref{CMsol}  enjoying axial symmetry. The fields read 
\begin{equation}
\label{sol1}
\cV= -\frac{\rho^2}{l^2}+ \frac{\cV_1}{\rho}\,,\quad 
\cU= \frac{\gamma l  \cE^{-3} }{ \rho} \,,    \quad 
\Phi=2 \log\left(\frac{\rho}{l} \cE\right)\,.
\end{equation}
where $\cV_1$ and $\cE$ are integration constants. This is a rotating spacetime as $\cU \neq 0$. Interestingly, the rotation cannot be brought to zero by means of a coordinate transformation. This, for instance, differs from a rotating BTZ black hole, where the angular momentum can be obtained from a boost in the $u-\phi$ plane, \cite{Martinez:1999qi}. 

The rest of the constant $\gamma>0$ and $l$ are fixed and define the self-interacting potential
\be
\label{V1}
V(\Phi)=-\frac{6}{l^2} e^{-\Phi}+\frac{3 \gamma^2}{2 l^4} e^{-4\Phi}\,.
\ee
Let us comment on some attributes of this potential. It is direct to see $V(\Phi)$ goes to zero exponentially fast for positive $\Phi$. This yields a sector with vanishing cosmological constant, allowing for the asymptotically locally flat solution \eqref{sol1} to arise at large values of the scalar field.  The potential also has a global minimum at $\Phi_{\rm c}=\tfrac{2}{3} \log\left(\tfrac{\gamma}{l} \right)$. At that point,   the potential becomes an effective cosmological constant 
\be
V(\Phi_{\rm c})=-\tfrac{9}{2l^2}\left(\tfrac{l}{\gamma} \right)^{2/3}\equiv-\frac{2}{L^2}\,.
\ee
Around this critical value the scalar field adquires a mass $m^2=\frac{8}{L^2}$ and the geometry becomes a locally AdS$_3$ spacetime. Using the standard Klein-Gordon inner product, the scalar field is normalizable there. In the context of AdS/CFT this means that  its subleading term, the ``VEV" is excited around this point. At the end of this section we comment on how this AdS$_3$ background emerges as a near-horizon limit of \eqref{sol1}, hence this critical points plays a role similar to that of the effective potential of the attractor mechanism.

There is a more general phase space than the one spanned by the solution presented here. Actually, it is possible to see that replacing $\gamma$ in terms of $L$ and $l$ in the metric and shifting the scalar field there are three free parameters on the solution.  In our boundary conditions for the metric, we have fixed $l$, ending up with two integration constants, one associated to the mass and other to the angular momentum, as we will see below.

\subsubsection*{Charges and first law of thermodynamics} 
The line element described by \eqref{sol1} represents a rotating black hole, whose event horizon is located at
\be
\rho_0^3=\frac{l^2}{2} \left(\cV_1 + \sqrt{\cV^2_1-4\gamma^2/ \cE^6 }\right) 
\ee
provided $\cV_1\geq 2 \gamma /\cE^3 $. This is a null  surface generated by a Killing horizon $\chi=-\d_u+\omega_H \d_\varphi$. The value of the angular potential $\omega_{H}$ and the surface gravity $\kappa$ follows from $\chi^{\mu} \nabla_\mu \chi^\nu=\kappa \chi^\nu$ when evaluated on $\rho=\rho_0$, 
\be
\label{potentials}
 \kappa=\frac{3}{2} \left( \frac{\rho_0}{l^2} -\frac{ l^2 \gamma^2}{\cE^6 \rho_0^5}  \right)\, \quad \omega_{H}= \frac{ l \gamma}{(\cE \rho_0)^3}\,.
\ee
The global charges of this geometry can be obtained from \eqref{chargesphig}. They are the mass, associated to $f=1$, and the angular momentum which is conjugated to $Y=1$. Their values reduce to
\be
\slashed{\delta} Q(\d_u) = \frac{1}{8 G l} \left( 2  \cE \delta \cV_1+3   \delta \cE \cV_1 \right)\,,  \quad Q(\d_\vp) = \frac{3}{8G }  \alpha \cE^{-2} \,.
\ee
It is worth emphasising that  the first law of thermodynamics still holds in this type of solution regardless of the non-integrability of the mass. Indeed using potentials \eqref{potentials} one finds that
\be
\slashed{\delta} Q(\d_u) = T_H \delta S + \omega_{H} \delta Q(\d_\vp)\,.
\ee
where $T_H=\frac{\kappa}{2\pi}$ is the Hawking temperature and $S$ corresponds to the black hole entropy. 

\subsubsection*{Extremal solution}
Let us finally comment on the extremal black hole which saturates the relation between the integration constants, $\cV_1= 2 \gamma\cE^{-3} $. The scalar field is the same as in \eqref{sol1}, but the line element becomes
\be
ds^2= \frac{\rho^2}{l^2} \cE^2 \left[-\left(\frac{\rho}{l}-\frac{l\cV_1}{2\rho^2}\right)^2du^2-2du\,d\rho+\rho^2 \left(d\vp + \frac{l\cV_1}{2\rho^3} du\right)^2\right]\,.
\ee
The horizon is given by the  value where the scalar potential $V(\Phi)$ reaches its minimum, $\rho^3_{\rm c}=l^2 \gamma \cE^{-3}  $. By zooming in around this region we will find that the near-horizon description is captured by the Coussaert-Henneaux self-dual geometry \cite{Coussaert:1994tu}, which is one-parameter family of solutions. To show this, let us perform the transformation 
\be
\rho^3=\rho_{\rm c}^3+ \epsilon \Delta\,, \quad u=\frac{U}{\epsilon}\, \quad \vp=\phi-  \frac{U}{l\epsilon}\,. 
\ee 
After taking the limit $\epsilon \to 0$ we find a one-parameter solution given by
\be
ds^2= \frac{1}{\rho_c^2}\left(\frac{\gamma}{l}\right)^{2/3}  \left[-\frac{\Delta^2}{l^2 \rho^2_{\rm c}}dU^2-\frac{2}{3}dU\,d\Delta+\rho_{\rm c}^4 \left(d\phi - \frac{\Delta }{l \rho_c^3} dU\right)^2\right]\,,\;\; \Phi=\frac{2}{3} \log\left(\frac{\gamma}{l} \right)\,,
\ee
which depended on one integration constant $\rho_c$, 
 and it satisfies $R^{\mu}_{\, \nu}=- \tfrac{9}{2l^2}\left(\tfrac{l}{\gamma} \right)^{2/3}\delta^{\mu}_{\, \nu}$  consistent with the minimum of the scalar potential. This is not a surprise, as it is known that the Coussaert-Henneaux geometry arises as a near-horizon limit of an extremal BTZ black hole \cite{Balasubramanian:2009bg}.

\subsection{Dynamical domain-wall}
\label{exact:2}

Another interesting solution in the gauge \eqref{CBondi} can be found for the self-interacting potential
\be
\label{Vk}
V(\Phi)=-\frac{6}{l^2} e^{-\Phi}+\frac{4k}{l^2}e^{-\frac{3}{2}\Phi}\,.
\ee
This potential shares the same features of the previous section. It has a minimum, where an AdS$_3$ space emerges and for large values of the scalar field the asymptotically locally flat spaces appear. In what follows, we will present a dynamical configuration that interpolates between those two vacua.
\begin{center}
\label{fig1}
\includegraphics[width=.9\linewidth]{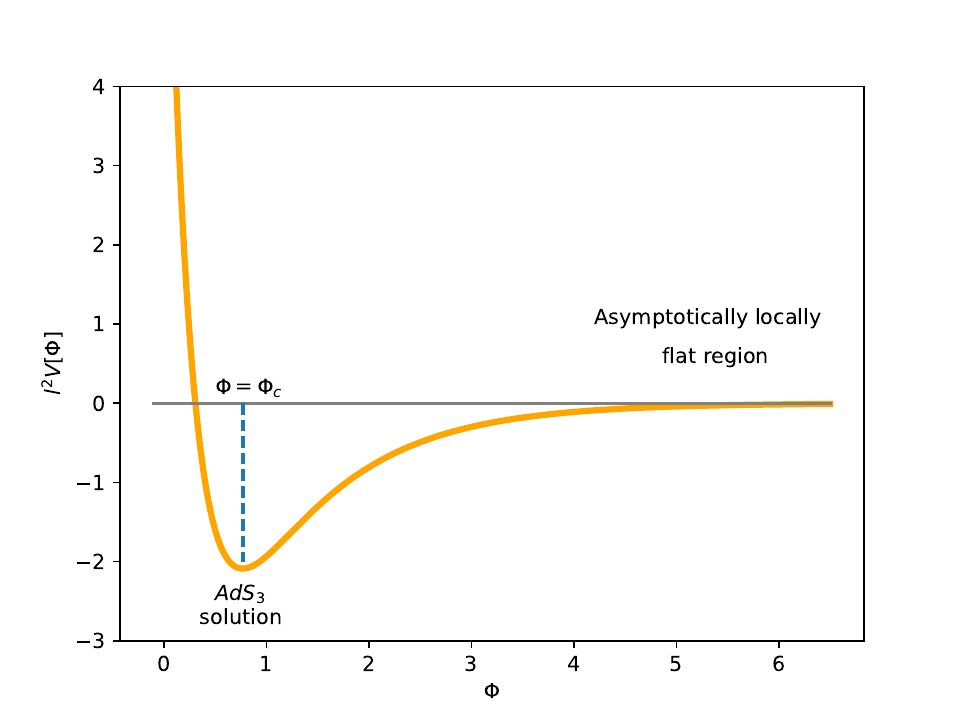}
\captionof{figure}{A schematic representation of the potential $l^2 V(\Phi)$ for $k=10$. At the minimum $\Phi_c=2\log(k)$ a locally AdS$_{3}$ solution arises representing, for instance, a BTZ black hole for the choice $\theta=\mu u$. For large values of $\Phi$, the curvature vanishes signaling the emergence of an asymptotically locally flat geometry. Notice that a very similar profile is also displayed by the scalar potential defined in \eqref{V1}.}
\end{center}
The scalar field takes the form
\begin{equation}
\label{Phiu}
\Phi=2 \log\left( \frac{\rho}{l}\cE-l\dot{\cE} + k \right)\,, 
\end{equation}
where now it is more convenient to explicitly use that $\cE=e^{-\theta(u,\vp)}$. This because $\theta$ satisfies the free boson equation
\begin{equation}
\label{fbeq}
\Box \theta=0.
\end{equation}
whose general solution is
\be
\label{sol}
\theta=\mu  \, u + \sigma \, \vp + \theta_+(x^+)+\theta_-(x^-)\,,
\ee
where we have explicitly separated the zero mode contributions associated to constants $\mu$ and $\sigma$ from the rest of the oscillators $\theta_\pm$ that are arbitrary functions of $x^\pm=u/l \pm \vp$. 

In terms of the free field $\theta$, one completely solves the metric functions where relations \eqref{V} and \eqref{U} reduce to 
\begin{equation}
\label{eq:VU}   
\mathcal{V}=-\frac{\rho^2}{l^2}+l^2 \dot\theta^2+ \theta'^2-2 \theta'' \,, \quad \mathcal{U}=l^2(\dot{\theta} \theta' -\dot{\theta}')\,.
\end{equation}
It is worthwhile noticing that one can recognize the values of $\cV_0=l^2 \dot\theta^2+ \theta'^2-2 \theta''$ as the energy density and $\cU_0=l^2(\dot{\theta} \theta' -\dot{\theta}')$ as the momentum density of the two-dimensional free boson $\theta$. This means that the phase space of this two-dimensional system is controlled by the coadjoint representation of the Virasoro group \cite{Witten:1987ty} in terms of the representatives $(\cU_0,\cV_0)$. There is an additional restriction, namely that the scalar field $\Phi$ must be single-valued on the circle, which in turn implies the condition 
\be
\theta(u,\vp+2\pi)=\theta(u,\vp)\,. 
\ee
In terms of the mode expansion, the latter implies that $\sigma=0$  and that $\theta_\pm$ are periodically well-defined functions. Furthermore, the momentum density $\cU_0$ must have a vanishing zero mode.

The linear dependence on $u$ makes the scalar field becomes constant at late Bondi time,  $\Phi(u \to \infty)=2 \log(k)$\footnote{Without loss of generality, we have set $\mu>0$. Negative values of $\mu$ only revert the behavior of the $\Phi$ along $u$. }. This value produces a  spacetime that is locally AdS$_3$, which in the Bondi gauge, has been previously reported in \cite{Barnich:2012aw}. The effective cosmological constant is $\Lambda_{\rm eff}=-\frac{1}{k^2 l^2}$ corresponding to the value of the scalar when the potential reaches its minimum.  

For simplicity, let us examine in more detail the zero mode solution $\theta=\mu u$. Computing the scalar curvature in terms of $\Phi$ we can summarize some of the most relevant features exhibited by the solution\footnote{A similar behavior is observed after analyzing the Kretschmann invariant.}
\begin{equation}
R=-\frac{2}{l^2}(8 e^{-\Phi} - 4 e^{-\tfrac{3}{2}\Phi} k - k^2 e^{-2\Phi})\,,
\end{equation}
where we can see that for $\Phi \to \infty$ the spacetime becomes asymptotically flat. Besides, it becomes constant $R \to 6\Lambda_{\rm eff}$ when the scalar goes to $2\log(k)$. In these coordinates, there this a curvature singularity when $\Phi \to -\infty$ or $\rho_{\rm sing}=-l^2 \mu -k l e^{-\mu u}$. Fortunately, this singularity is hidden behind the null surface $\rho_{\rm null}=l^2 \mu$ for $k>0$. Therefore, the zero mode geometry of this solution represents a dynamical collapsing spacetime with time-dependent conformal factor that settles down to a non-rotating BTZ black hole at $u \to +\infty$. This final state corresponds to the minimum of the potential. 

An analysis of this thermalization behavior at the level of the integrable charges and its corresponding algebra is shown in the next section. 

\section{Charge algebra}
\label{alg}
Let us recall the expression for surfaces charges  found in section \eqref{sc}
\be
\label{charges21}
\slashed{\delta} Q_\xi (\cX) =\delta \cQ(s,\cX) +\Theta_s(\cX,\delta \cX)\,,
\ee
which for solutions \eqref{eq:VU} reduces to 
\begin{equation}
\label{charge22}
\begin{split}
\cQ(s,\cX)&=\frac{1}{16 \pi G} \oint \, (k-l\dot{\cE})\left( f \cV_0 + 2Y\cU_0\right),\\
\Theta_s(\cX,\delta \cX)&=\frac{1}{16 \pi l G} \oint \, \left(4 f'  \cU_0 \delta \cE + 2 f \cU_0' \delta \cE + l^2 f \cV_0 \delta \dot{\cE}  \right)\,.
\end{split}
\end{equation} 
In the standard construction \cite{Brown:1986ed,Brown:1986nw,Barnich:2001jy}, the algebra of the charges is isomorphic to the Lie bracket of the symmetries. This is possible only when ${\delta}Q_\xi$ is an exact $\delta$-form. In that case, the Dirac bracket of the charges is given by
\be
\{Q_{\xi_1},Q_{\xi_2}\}^{*}\equiv \delta_{\xi_2} Q_{\xi_1}
\ee
We would still like to have a canonical representation of the symmetries, however we cannot use the standard Dirac bracket due to the presence of $\Theta_s(\cX,\delta \cX)$ in the charges. Fortunately, Barnich and Troessaert (BT) \cite{Barnich:2011mi} have constructed a bracket for the integrable charges $\cQ(s,\cX)$ that incorporates the non-integrable contribution. The mentioned algebraic structure is given by
\be
\label{BT}
\{\cQ(s_1,\cX),\cQ(s_2,\cX)\}_{\scalebox{.6}{\rm BT}}\equiv \delta_{s_2} \cQ(s_1,\cX) + \Theta_{s_2}(\cX,\delta_{s_1} \cX)\,,
\ee
At this point, we need the action of the asymptotic vectors $s\equiv (f,Y)$ on the leading fields. The relevant field is just $\theta$ and its transformation law can be obtained from the preservation of the fall-off at infinity \eqref{asym}. Doing so, we get 
\be
\label{trthe}
\delta_s \theta= f \dot{\theta}+ Y \theta'+Y'\,.
\ee
The transformation laws of fields $\cV_0$ and $\cU_0$ can be obtained from the previous relations. They yield
\begin{equation}
\label{transT}
\begin{split}
\delta_s \cV_0&= Y \cV_0'+2Y'\cV_0-2Y''' +\frac{2}{l^2}(f \cU_0' +2 f' \cU_0 )\,,\\
\delta_s \cU_0&=Y \cU_0' +2 Y' \cU_0 + \tfrac{1}{2}f\cV_0' +f' \cV_0 -f''' \,.   
\end{split}
\end{equation}
By using these relations we can show that after an straightforward, albeit lengthy  application of the BT bracket formula, gives a canonical representation of the asymptotic symmetries. Indeed, by collecting all terms and removing the dependence on $\cX$ for simplicity, we find
\begin{equation}
\label{resBT}
\{\cQ(s_1),\cQ(s_2)\}_{\scalebox{.6}{\rm BT}}=\cQ([s_1,s_2])+ \cK(s_1,s_2)\,  
\end{equation}
where $[s_1,s_2]$ is the modified Lie bracket of two Killing vectors \eqref{Lie12} and $\cK(s_1,s_2)$ is \emph{field dependent} central extension 
\begin{equation}
\label{KF0}
\cK(s_1,s_2)=\frac{1}{8\pi G} \oint \Big[ (k-l\dot{\cE})(Y_1''' f_2-f_1 Y_2''')+ \frac{1}{l}\cE''' (f'_1 f_2  - f_1 f_2'  ) \Big]\,.
\end{equation}
This term is a new element of the algebra generated by the BT bracket, provided
\be
\label{resBT2}
\{\cK(s_1,s_2),\cQ(s_3)\}_{\scalebox{.6}{\rm BT}}\equiv \delta_{s_3} \cK(s_1,s_2)\,, \quad   \{\cK(s_1,s_2),\cK(s_3,s_4)\}_{\scalebox{.6}{\rm BT}}=0\,.
\ee
It is important to stress that since $\cK_{s_1,s_2}$ transforms under the ASG, the Jacobi identity is satisfied by \eqref{resBT}, provided the field dependent central extension satisfies a generalized cocycle condition
\be
\label{cocycleG}
\delta_{s_3} \cK(s_1,s_2)+ \cK([s_1,s_2],s_3)+{\rm cyclic}(1,2,3)=0\,,
\ee
which is indeed satisfied by \eqref{KF0} (details can be found in appendix \ref{gcc}).

The field dependent central extension $\cK(s_1,s_2)$ has ambiguities associated to the normalization of the integrable part of the charge\cite{Barnich:2011mi}. In section \ref{sc}, we mention that $\cQ(s)$ can be shifted by a normalization $N_s$, so that the total non-integrable charge $\slashed{\delta} Q_{\xi}$ remains invariant. At the level of the algebra, we obtained \eqref{resBT} for the charges $\cQ'(s)=\cQ(s)+N_s$ and $\cK'(s_1,s_2)$ given by
\be
\cK'(s_1,s_2)=\cK(s_1,s_2)+\delta_{s_1}N_{s_2}-\delta_{s_1}N_{s_2}+N_{[s_1,s_2]} \,.
\ee
An ambiguity of this sort is considered trivial as the generalized cocycle condition is \eqref{cocycleG} immediately satisfied.

\subsection{Mode expansion}
It is enlightening to express the generators of the algebra \eqref{resBT} in terms of the Fourier modes solutions to \eqref{eq:fY} 
\be
\ell^\pm_{n}=\tfrac{1}{2}e^{i n \left(\tfrac{u}{l}\pm \vp\right) } (l \d_u \pm \d_\vp) \,, \quad \mathscr{L}^\pm_{n}=\cQ(\ell^\pm_{n})\,.
\ee
The algebra \eqref{resBT} can be explicitly written as
\begin{equation}
\label{KF}
\begin{split}
i\{\mathscr{L}^\pm_{n},\mathscr{L}^{\pm}_m\}_{\scalebox{.6}{\rm BT}}=(n-m) \mathscr{L}^{\pm}_{m+n}+\cK(\ell_n^\pm,\ell_m^\pm)\,, \quad 
i\{\mathscr{L}^+_{n},\mathscr{L}^-_m\}_{\scalebox{.6}{\rm BT}}=\cK(\ell_n^+,\ell_m^-)\,. 
\end{split}
\end{equation}
with
\begin{equation}
\label{AlgKF}
\begin{split}
\cK(\ell_n^\pm,\ell_m^\pm)&= \frac{kl}{8G} n^3 \delta_{m+n,0} - \frac{l}{16 G}e^{i(m+n)\tfrac{u}{l}}\left[(n^3-m^3) l\dot{\cE}_{m+n} + i (n-m)(m+n)^3 \cE_{m+n}\right]\\
\cK(\ell_n^+,\ell_m^-)&=\frac{l}{16 G} e^{i(m+n)\tfrac{u}{l}} \left[(n^3-m^3) l \dot{\cE}_{n-m} -i (n-m)^4 \cE_{n-m} \right]\,.
\end{split}
\end{equation}
This is an explicitly time-dependent algebra with $\cE_n(u)=\tfrac{1}{2\pi} \oint e^{in\vp} \, \cE(u,\vp)$. However, it has an interesting late time behavior since the coefficient $\cE_n$ are exponentially decaying for large $u$. 
This can be seen by using the periodic properties of $\theta$ defined in \eqref{sol}, implying
\be
\label{decaying}
\cE_n(u)=e^{-\left(\mu-i\tfrac{n}{l}\right)u} \sum^{\infty}_{k=-\infty} \cE^+_{k} \cE^{-}_{n+k}e^{2 i\tfrac{k u}{l}}\,,
\ee
with $\cE^\pm_n$ being the Fourier mode of the functions $e^{-\theta_\pm}$. The important conclusion we get from the mode expansion of $\cE_n(u)$ is that only the term proportional to $k$ in \eqref{KF} survives in the limit $u\to+\infty$. Therefore, we have
\be
\cK(\ell_n^\pm,\ell_m^\pm) \to \frac{c}{12} m^3 \delta_{m+n,0}\,, \quad \cK(\ell_n^+,\ell_m^-)\to 0\,,
\ee
and the time-dependent algebra \eqref{AlgKF} becomes two commuting copies of the Virasoro algebra with
\be
c=\frac{3 k l}{2G}\,,
\ee
being the Brown-Henneaux central charge. This asymptotic symmetry algebra emerges for configurations that slightly depart from the  minimum of the potential \eqref{Vk}.

\section{Discussion}
\label{s7}

In this work we have analyzed the asymptotic structure of Einstein gravity coupled to a self-interacting scalar field in $2+1$ dimensions. We have focused on the sector in which the self-interaction permits a scalar field with slow fall-off at infinity, compatible with asymptotically locally flat geometries. Remarkably, in spite of the slow fall-off of the scalar and the non-trivial asymptotic geometry, the charges turn out to be finite due to a subtle cancellation of divergences coming from both the gravity and the matter sectors. The potentials here considered are of the same form as those emerging from Kaluza-Klein compactifications of GR coupled to matter fields \cite{Carroll:2001ih} and in gauged supergravity (see e.g. \cite{Freedman:1978ra}, \cite{Gubser:2001eg} and \cite{OColgain:2010yxb}). We have proved that for a family of self-interactions, even in the asymptotically locally flat region, two copies of the Witt algebra emerge, since the behavior of the metric at infinity is such that our geometries share the conformal asymptotic geometry with AdS$_3$. By working on a conformal Bondi gauge, we have shown that the field dependent asymptotic symmetries span two copies of the Witt algebra, which is canonically realized by means of the Barnich-Troessaert bracket \cite{Barnich:2011mi}. Even more, we were able to find a new infinity family of time dependent solutions. These configurations are governed by a two-dimensional free-boson theory, which dynamically interpolate between the asymptotically locally flat and the locally AdS vacua of the theory, the latter being achieved for late times. The canonical realization of the charges associated to these dynamical solutions is given by two copies of the centrally extended Witt algebra, where the central extension matches the Brown-Hennaux one for late times. This can be interpreted as a realization of the Holographic $c-$theorem, since the spacetime reaches a locally AdS geometry for late times, and therefore the time plays the role of the holographic coordinate leading to a flow of the central charge.

In vacuum, the dynamics of GR is purely given in terms of boundary gravitons, while the inclusion of a scalar field introduces a local degree of freedom which, as we have shown, suitably modify the rich asymptotic structure of GR in vacuum. We have proposed a family of asymptotic behaviors for the metric and the scalar field in the conformal Bondi gauge, which accommodates black holes that were already known in the literature \cite{Chan:1994qa,Chan:1995wj} as well as new stationary and even dynamical solutions. We have provided a consistent thermodynamics for the new stationary black holes, which are characterized by two integration constants, and lead to a mass that, in spite of being non-integrable, fulfills the first law of black hole thermodynamics. It would be interesting to study the structure of phase transitions that may emerge within these solutions and the thermal background, or even between the solutions in the AdS sector of the theories we have considered, and the BTZ solution that is achieved for a fixed value of the scalar field, sitting at a negative minimum of the self-interacting potential. We expect to report along these lines in the near future.


\subsection{Comparison to the Jordan frame}
\label{cjf}
The problem studied in this work can be re-expressed in a different conformal frame by means of the transformation \be
\label{eq:1}
g_{\mu \nu}=e^{\Phi} \bar{g}_{\mu \nu}\,,
\ee
consistent with boundary condition presented in \eqref{CBondi}. 
Doing so, action \eqref{eq:action} is transformed (up to a boundary term) to
\begin{equation}
\label{eq:2}
S_{0}[\bar{g}_{\mu \nu},\Phi]= \frac{1}{16\pi G}\int d^3x \sqrt{-\bar{g}} e^{\tfrac{\Phi}{2}}\left[\bar{R} - e^{\Phi} V(\Phi) \right]\,,
\end{equation}
where now the kinetic potential is now absent and the interaction of the scalar with the geometry is due to the interaction with scalar curvature. For the space of solutions studied in section \eqref{ASYMP}, we find that
\be
e^{\Phi} V(\Phi)=-\frac{6}{l^2} + \frac{4k}{l^2} e^{-\frac{\Phi}{2}}+O(e^{-\frac{3\Phi}{2}})
\ee
It is interesting to note that this term is a self-interacting deformation of the constant term appearing in the low energy action arising from string theory (One can compare with \cite{Kanitscheider:2008kd} after setting $\beta=0$ in action (3.4) of that article).

One can also translate the asymptotic structure of the configurations studied in this article to the ones controlled by \eqref{eq:2}. In terms of the conformally transformed metric \eqref{eq:1}  boundary conditions studied in section \eqref{s3} can be expressed using the same gauge fixing yielding
\begin{equation}
\label{CBondi2}
    d\bar{s}^2= \mathcal{V} du^2 - 2 du d\rho + 2\,\mathcal{U} du d\varphi + \rho^2 d\varphi^2\,,
\end{equation}
and with the same profile for the scalar field. Then, the analog of boundary conditions \eqref{gfc} and \eqref{asym} become
\be
\label{gfc2}
\begin{split}
\mathcal{L}_\xi \bar{g}_{\rho\rho}=\mathcal{L}_\xi \bar{g}_{\rho\varphi}&= \mathcal{L}_\xi \bar{g}_{\varphi\varphi}=0\,,\\
\mathcal{L}_\xi \bar{g}_{u u}=\mathcal{L}_\xi\bar{g}_{u\varphi}=O(1)\, \quad &\mathcal{L}_\xi \bar{g}_{u\rho}=0 \,, \quad \mathcal{\cL}_\xi \Phi=O(1)\,.
\end{split}
\ee
The metric field exhibits Brown-Henneaux boundary conditions at $\rho\to \infty$ expressed in the BMS gauge \cite{Barnich:2012aw} together with the presence of a non-vanishing scalar field at the boundary. Moreover, it is important to point out that \eqref{gfc2} also corresponds to the asymptotic behavior discussed in section 5.1 of \cite{Kanitscheider:2008kd} when working in the Fefferman-Graham (FG) gauge for the particular case of $d=2$ spatial dimensions at the boundary. Since the asymptotic behavior displayed by the metric $\bar{g}_{\mu\nu}$ is the same than in Einstein gravity, it has been concluded in \cite{Barnich:2012aw} that BMS and FG gauges contain the same physical information.

Finally, one must compute the expression for the surface charges associated to the model \eqref{eq:2}. However, we can take a shortcut in the derivation of the quantities by taking into account the covariant form of \eqref{chargesphig0}. In fact, one can readily yield an expression for them using the conformal transformation \eqref{eq:1} and by expressing everything in terms of the functional variation
\be
\delta g_{\mu \nu}= (\delta \bar{g}_{\mu \nu}+\delta \Phi \bar{g}_{\mu \nu}) e^{\Phi}\,, 
\ee
along with its corresponding covariant derivative $\bar{\nabla}$. This procedure, of course, will lead to the same charges computed in section \ref{s4}.


\section*{Acknowledgments}

We thank  Gaston Giribet, Olivera Mišković Alfredo P\'erez, Francisco Rojas  and Ricardo Troncoso for useful discussions. This research has been supported by FONDECYT grants 1200986, 1210635, 1221504, 1221920 and 1230853. The research of AA is supported in part by a visiting researcher award of the FAPESP 2022/11765-7. AN thanks Beca ANID-Subdirección de Capital Humano/Magíster Nacional/2021-22211733 for financial support. HG and AN would also like to acknowledge the support of ACT210100 ANILLO Grant.

\appendix
\section{Exact solution for arbitrary $q$}
\label{Exactq}
One can find an exact and dynamical solution to the Einstein-scalar system where the parameter $q$ (discussed in section \ref{s2}) is an arbitrary real number. It has a similar structure than the solution studied in \eqref{exact:2} and is given by
\begin{equation}
\label{CBondi2}
ds^2=e^{\sqrt{\frac{2q}{q+1}}\Phi} \left[ \mathcal{V} du^2 - 2 du d\rho + 2\,\mathcal{U} du d\varphi + \rho^2 d\varphi^2 \right]\,,\\
\quad \Phi=\sqrt{2q(q+1)}\log\left( \frac{\rho}{l}\cE-l\dot{\cE} + k \right)\,.
\end{equation}
where $\cV$ and $\cU$ are given in \eqref{eq:VU} and $\cE=e^{-\theta}$ is again determined by the free wave equation \eqref{fbeq}. Apart from the different asymptotic structure displayed by \eqref{CBondi2}, another important difference is that the potential $V(\Phi)$ is now 
\begin{equation}
V(\Phi)=-\frac{1}{l^2}\left(k^2 q(q-1)e^{-\sqrt{2\left(1+\frac{1}{q} \right)}\Phi}-2k q(q+1) e^{-\frac{2q+1}{\sqrt{2q(q+1)}}\Phi} + (q+1)(q+2)e^{-\sqrt{\frac{2q}{q+1}} \Phi} \right)\,.
\end{equation}
Interestingly, the potential  studied in \eqref{Vk} is easily recovered for $q=1$. Furthermore, it has now two extrema. One of them is the maximum at $\Phi_{\rm max}=\sqrt{2q(q+1)}\log\left(k\tfrac{q-1}{q+2}\right)$ that arises for $q>1$ and represents an unstable dS$_3$ solution. The other critical point is a local minimum at $\Phi_{\rm min}=\sqrt{2q(q+1)}\log\left(k\right)$ where again one has a negative cosmological constant and the maximally symmetric solution is AdS$_3$ there. This vacuum will be a local minimum of the energy provided the potential comes from a globally well-defined superpotential. This is indeed the case as
\begin{equation}
  V(\Phi)=2\left( \frac{dP(\Phi)}{d\Phi} \right)^2-2P(\Phi)^2\,,
\end{equation}
where the function $P(\Phi)$ is 
\begin{equation}
    P(\Phi)=k\, q\, e^{-\sqrt{\frac{1}{2}\left(1+\frac{1}{q}\right)}\Phi} - \frac{1+q}{l} e^{-\sqrt{\frac{q}{2(q+1)}}\Phi}\,.
\end{equation}

\section{Generalized cocycle condition}
\label{gcc}
Here we show that 
\be
\label{Gcocycle}
\delta_{s_3} \cK(s_1,s_2)+ \cK([s_1,s_2],s_3)+{\rm cyc}(1,2,3)=0\,
\ee
is satisfied for \eqref{KF}. To do so, we define $\cK(s_1,s_2)\equiv \cA_{12}+\cB_{12}$ such that
\begin{equation}
\cA_{12}=\frac{1}{8\pi G} \oint (k-l\dot{\cE})(Y_1''' f_2-f_1 Y_2''')\, \quad \cB_{12}=\frac{1}{8\pi G l} \oint \cE''' (f'_1 f_2  - f_1 f_2')
\end{equation}
Then
\begin{equation}
\delta_3 \cA_{12}+\rm {cyc}(1,2,3)=\frac{1}{8\pi G} \oint \left[\tfrac{1}{l} \left( f''_3\cE-f'_3\cE' \right) - l Y_3 \dot{\cE}'\right](Y_1''' f_2-f_1 Y_2''')+{\rm cyc}(1,2,3)
\end{equation}
and
\begin{equation}
\begin{split}
\cA_{[1,2]3}+{\rm cyc}(1,2,3)&=\frac{1}{8\pi G} \oint (-l\dot{\cE})(Y_{[1,2]}''' f_3-f_{[1,2]} Y_3''')+(1,2,3)\\
&=\frac{1}{8\pi G} \oint (-l \dot{\cE}) \Big[ Y_3( Y_1''' f_2- f_1 Y_2''') + \frac{2}{l^2}f_3(f_1' f_2''-f_1'' f_2') \Big]'+{\rm cyc}(1,2,3)\,,
\end{split}
\end{equation}
thus,
\begin{multline}
\label{1}
\delta_3 \cA_{12}+\cA_{[1,2]3}+{\rm cyc}(1,2,3)\\
=\frac{1}{8\pi G l} \oint \left[ \left(f''_3\cE-f'_3\cE'\right) (Y_1''' f_2-f_1 Y_2''')-2 \dot{\cE} f_3(f_1' f_2'''-f_1''' f_2')\right]+(1,2,3)\,.
\end{multline}
Analogously, for $\cB_{12}$ we have
\begin{multline}
\label{2}
\delta_3 \cB_{12}+{\rm cyc}(1,2,3)=\frac{1}{8\pi G l} \oint \Big[-2f_3 \dot{\cE}(f_1''' f_2'-f_1' f_2''')\\-(Y_3'\cE''-Y_3'''\cE+Y_3\cE'''-Y_3''\cE')(f_1'' f_2-f_1 f_2'') \Big]+{\rm cyc}(1,2,3)
\end{multline}
and 
\begin{equation}
\begin{split}
\cB_{[1,2]3}+{\rm cyc}(1,2,3)&=\frac{1}{8\pi Gl} \oint \cE'''\Big[f_3(f_1'Y_2'-Y_1'f_2'+Y_1 f_2''-Y_2 f_1'')\Big]+{\rm cyc}(1,2,3)\,.
\end{split}
\end{equation}
that can be used to obtain
\begin{multline}
\label{3}
\delta_3 \cB_{12}+\cB_{[1,2]3}+{\rm cyc}(1,2,3)=\frac{1}{8\pi Gl} \oint\Big[-2f_3 \dot{\cE}(f_1''' f_2'-f_1' f_2''') +Y_3'''\cE(f_1'' f_2-f_1 f_2'')\\ +\cE'f_3(f_1'Y_2'''-Y_1'''f_2') \Big]+{\rm cyc}(1,2,3)\,.
\end{multline}
By summing up contributions \eqref{1} and \eqref{3} together with their cyclic permutations, one gets the desired result \eqref{Gcocycle}.


\end{document}